%% file: prl_Kt_collab3_1.tex
\documentclass[aps,prl,twocolumn,superscriptaddress,letterpaper]{revtex4}

\usepackage{graphicx}
\begin{document}

% Use the \preprint command to place your local institutional report
% number in the upper righthand corner of the title page in preprint mode.
% Multiple \preprint commands are allowed.
% Use the 'preprintnumbers' class option to override journal defaults
% to display numbers if necessary
%\preprint{}

%Title of paper
\title{Measurement of the $k_{T}$ Distribution of Particles in Jets Produced in $p\bar p$ Collisions at $\sqrt{s}=1.96$ TeV}

% repeat the \author .. \affiliation  etc. as needed
% \email, \thanks, \homepage, \altaffiliation all apply to the current
% author. Explanatory text should go in the []'s, actual e-mail
% address or url should go in the {}'s for \email and \homepage.
% Please use the appropriate macro foreach each type of information

% \affiliation command applies to all authors since the last
% \affiliation command. The \affiliation command should follow the
% other information
% \affiliation can be followed by \email, \homepage, \thanks as well.
\author{}
%\email[]{Your e-mail address}
%\homepage[]{Your web page}
%\thanks{}
%\altaffiliation{}
\affiliation{}

%Collaboration name if desired (requires use of superscriptaddress
%option in \documentclass). \noaffiliation is required (may also be
%used with the \author command).
%\collaboration can be followed by \email, \homepage, \thanks as well.
%\collaboration{}
%\noaffiliation

\date{\today}

\begin{abstract}
We present a measurement of the transverse momentum with respect to the jet axis ($k_{T}$) of particles in jets produced in $p\bar p$ collisions at $\sqrt{s}=1.96$ TeV. Results are obtained for charged particles within a cone of opening angle 0.5 radians around the jet axis in events with dijet invariant masses between 66 and 737 GeV/c$^{2}$. The experimental data are compared to theoretical predictions obtained for fragmentation partons within the framework of resummed perturbative QCD using the modified leading log and next-to-modified leading log approximations. The comparison shows that trends in data are successfully described by the theoretical predictions, indicating that the perturbative QCD stage of jet fragmentation is dominant in shaping basic jet characteristics. 
\end{abstract}

% insert suggested PACS numbers in braces on next line
\pacs{}
% insert suggested keywords - APS authors don't need to do this
%\keywords{}

%\maketitle must follow title, authors, abstract, \pacs, and \keywords
\input{July_2008_Authors_Visitors}
\maketitle
	
	 In this analysis we measure the transverse momenta of particles in jets with respect to the jet axis ($k_{T}$), study the dependence of the $k_T$ distribution on jet energy, and compare the results to analytical predictions of the modified leading log approximation (MLLA) {\cite{RedamyKtMLLA}} and next-to-modified leading log approximation (NMLLA) {\cite{RedamyKtNMLLA}}, supplemented with the hypothesis of local parton-hadron duality (LPHD) {\cite{lphd}}. 

	This measurement tests the applicability of perturbative QCD (pQCD) to the soft process of jet fragmentation.  Detailed studies of jet fragmentation expand our understanding of the relative roles of the perturbative and non-perturbative stages of jet formation, and they probe the boundary between the parton shower and hadronization.  The ultimate goal is to understand which stage of jet formation is most significant in determining the final characteristics of jets.  This measurement indicates that the parton shower dominates.  Moreover, we also verify how well the {\sc pythia} tune A {\cite{Pythia,PythiaTuneA1}} and {\sc herwig} 6.5 {\cite{Herwig}} Monte Carlo generators describe jet properties in the data.  This comparison is crucial for data analyses utilizing these generators, and the results can be used to tune the generators for future measurements.
	
	Past experimental studies of the inclusive distributions of particles in jets {\cite{Gary,SafonovPRL,SafonovPRD}} and the recent measurement of the two-particle momentum correlation in jets {\cite{Correlations}} agree well with theoretical predictions, suggesting that the perturbative QCD stage of jet formation is dominant. In this analysis, we further test the LPHD hypothesis by examining whether the pQCD predictions for the transverse momentum distribution of partons can successfully reproduce the corresponding distribution for hadrons in experimental data. We report a measurement of the $k_{T}$ distribution $dN/d(\ln k_{T})$ of charged particles in restricted cones with an opening angle of $\theta_{c}=0.5$ radians around the jet axis in events with dijet invariant masses in the range 66--737 GeV/c$^{2}$. It has been shown in the past that the integral of the distribution is well described by MLLA predictions {\cite{SafonovPRD}}; therefore, in this article we compare only the shape of the distribution by normalizing theory to data in the region $-0.2<\ln[k_{T}/$(GeV/c)$]<0.0$, where both the theoretical prediction and the experimental measurement are expected to be reliable. The data are corrected for detector effects and no additional corrections are needed for comparison to the theoretical predictions if LPHD is assumed.
	
	The theoretical predictions used in this analysis are formulated for dijet events. MLLA {\cite{MLLAgeneral}} is an approximation which allows one to calculate a variety of observables via a complete resummation of perturbative terms. It is an approximation in the sense that each perturbative term of order $n$ is calculated to a precision of leading and next-to-leading logarithms:
\begin{eqnarray}
\alpha_{s}^{n}(k_{T})( A_n \ln^{2n}(E_\mathit{jet}) + B_n \ln^{2n-1}(E_\mathit{jet}) \\
 +~O( \ln^{2n-2}(E_\mathit{jet}))) \nonumber,
\end{eqnarray}
where $\alpha_{s}(k_{T})$ is the strong coupling constant. The constants $A_n$ and $B_n$ are calculated exactly to all orders. The NMLLA calculations extend the MLLA precision by treating a number of contributions more consistently at the next-to-MLLA level, i.e. at the level of
$\alpha_{s}^{n}(k_{T})\ln^{2n-2}(E_\mathit{jet})$. Therefore, MLLA and NMLLA both provide soft gluon resummation, but at different levels of precision.
	
	The MLLA + LPHD and NMLLA + LPHD approaches view jet fragmentation as a predominantly perturbative QCD process. The MLLA and NMLLA calculations predict the average number of partons $N$ and the transverse momentum distribution of partons with respect to the direction of the initial parton. The predictions are valid for partons in a small cone with opening angle $\theta_{c}$ around the direction of the initial parton and they assume that the parton momentum is much smaller than the jet energy (soft approximation). The predictions are functions of $Y=\ln (Q/Q_\mathit{eff})$, where $Q=E_\mathit{jet}\theta_{c}$ is the so-called jet hardness and $Q_\mathit{eff}$ is the lowest allowed transverse momentum of partons. The LPHD hypothesis states that the hadronization process takes place locally and, therefore, properties of partons and hadrons are closely related. For instance, the parton and hadron $k_{T}$ distributions are assumed to be related via a constant factor $K_{LPHD}$, which is independent of the jet energy and whether the jet originates from a quark or a gluon {\cite{KhozeKLPHD}}. Past studies have shown that $K_{LPHD}\sim 1$ {\cite{SafonovPRD}}.

\begin{figure}[h!]
	\includegraphics[width=2.52in]
	{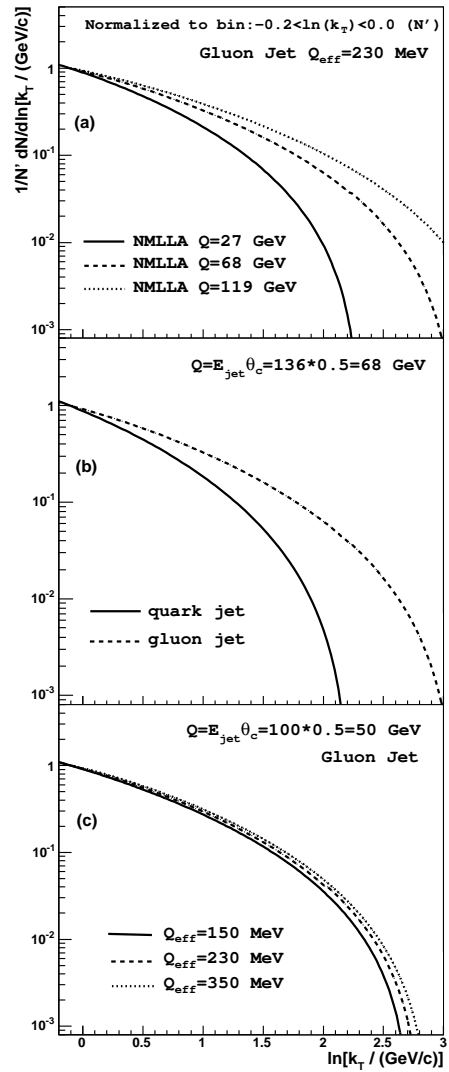}
	\caption{NMLLA predictions {\cite{RedamyKtNMLLA}} of the $k_{T}$ distribution in jets. The figures show how the distribution depends on (a) the jet hardness (shown for a gluon jet), (b) the origin of the jet (quark or gluon), and (c) the parton shower cutoff $Q_\mathit{eff}$.}
	\label{PredictTheory}
	\end{figure}

	Theoretical NMLLA predictions for the $k_{T}$ distribution are shown in Fig.~{\ref{PredictTheory}}. The direction of the initial parton is used as the jet axis. The lower boundary of the range of validity of the predictions is determined by $Q_\mathit{eff}$ and is $k_{T}> Q_\mathit{eff}$; however, in this measurement we only consider particles with $k_{T} > 0.5$ GeV/c ($\ln[k_{T}$/(GeV/c)] $>$ --0.6), motivated by the poor reconstruction quality of tracks with low $k_{T}$. The upper boundary is determined by the soft approximation requirement $k_{T}/E_\mathit{jet}\ll 1$ and the requirement that the double differential distribution ($\frac{d^{2}N}{dk~dk_{T}}$) be positive over the perturbative region {\cite{RedamyKtMLLA}}. This translates into $\ln[k_{T}$/(GeV/c)] $\lesssim  \ln(Q$/GeV)$- 2.5$ for MLLA and $\ln[k_{T}$/(GeV/c)] $\lesssim \ln(Q$/GeV)$ - 1.6$ for NMLLA  \cite{KtRange}. The range of validity extends to higher $k_{T}$ regions for increasing jet energy. The shape of the distribution shows a weak dependence on the value of $Q_\mathit{eff}$. 

    	Jets originating from gluons are expected to have more particles with large $k_{T}$, on average, than jets originating from quarks. In theory, the $k_{T}$ distribution is calculated for quark and gluon jets separately. Dijet events at the Tevatron consist of both quark and gluon jets. In order to compare data to theory, we rewrite the formula for the predictions as follows:
	\begin{eqnarray}
	\frac{dN}{d\ln(k_{T})}=f_{g}\left(\frac{dN}{d\ln(k_{T})}\right)_{g}+(1-f_{g})\left(\frac{dN}{d\ln(k_{T})}\right)_{q},
	\end{eqnarray}
	where $(\frac{dN}{d\ln(k_{T})})_{q}$ and $(\frac{dN}{d\ln(k_{T})})_{g}$ are the predictions for quark and gluon jets, respectively, and $f_{g}$ is the fraction of gluon jets in the data.

%	\begin{figure}[h!]
%	\includegraphics[width=2.55in]
%	{../Ktplots/mb5_NMLLA_Q_Log_mod2.eps}
%	\includegraphics[width=2.55in]
%	{../Ktplots/mb5_NMLLA_q_g_Log_mod2.eps}
%	\includegraphics[width=2.55in]
%	{../Ktplots/mb5_NMLLA_Qeff_Log_mod2.eps}
%	\caption{NMLLA predictions {\cite{RedamyKtNMLLA}} of the $k_{T}$ distribution in jets. The figures show (a) the dependence of the distribution for a gluon jet on the jet hardness, (b) the origin of the jet --- quark or gluon, and (c) the value of the parton shower cutoff $Q_\mathit{eff}$.}
%	\label{PredictTheory}
%	\end{figure}
	The measurement is based on events produced at the Tevatron collider in $p\bar p$ collisions at a center of mass energy of 1.96 TeV and recorded by the CDF II detector. The total integrated luminosity is 775 pb$^{-1}$. A detailed description of the CDF II detector can be found in {\cite{CDFII}} and references therein. Here we briefly describe the components of the detector that are relevant to this analysis. The silicon microstrip detector is used to reconstruct event vertices and to measure the distance of closest approach, $d_{0}$, of charged particles to the beamline in the plane transverse to the beam direction. The silicon detector is surrounded by the central outer tracker, an open-cell drift chamber providing up to 96 measurements of a charged particle track over the radial region from 40 to 137 cm. The entire CDF II tracking system is located inside a 1.4~T solenoidal magnet and is surrounded by calorimeters used to measure the energy of charged and neutral particles. The central electromagnetic, central hadronic, and wall hadronic calorimeters are made of lead (electromagnetic) and iron (hadronic) layers interspersed with plastic scintillator. The CDF II trigger system is a three-level filter with calorimeter information available at the first level {\cite{trigger}}.

	In this measurement, jets are reconstructed based on calorimeter information using a cone algorithm with cone radius $R=\sqrt{(\Delta\phi)^{2}+(\Delta\eta)^{2}}=1.0$ {\cite{JetClu}}. The energy of each jet is then corrected to compensate for the non-linearity and non-uniformity of the energy response of the calorimeter, the energy deposited inside the jet cone from sources other than the leading parton, and the leading parton energy deposited outside the jet cone. A detailed description of this procedure can be found in {\cite{JetCluCorr}}. The overall uncertainty on the jet energy scale is $3\%$. 

	In this article we give a brief overview of the event and track selection; a detailed description of the procedure and the evaluation of the systematic uncertainties can be found in {\cite{Correlations}}. Events were collected using a single calorimeter tower trigger with a transverse energy ($E_{T}$) \cite{Et} threshold of 5~GeV and with single jet triggers with $E_{T}$ thresholds of 20, 50, 70, and 100 GeV. To reject events with poorly measured jets, we require the two leading jets to be well balanced in $E_{T}$. We allow up to two extra jets, but their energy is required to be small: $E_{T}^\mathit{extra}<5.5$ GeV $+0.065(E_{T}^{1}+E_{T}^{2})$, where $E_{T}^{1}$, $E_{T}^{2}$, and $E_{T}^\mathit{extra}$ are the transverse energies of two leading jets and an extra jet, respectively. The final sample consists of approximately 250~000 events and is further divided into eight bins according to the dijet mass as measured by the calorimeters and defined as $M_{jj}=\sqrt{(E_{1}+E_{2})^{2}/c^{4}-(\vec{P}_{1}+\vec{P}_{2})^{2}/c^{2}}$, where $E$ and $\vec{P}$ are the energies and momenta of the two leading jets, respectively. Measurements are performed in the dijet center of mass frame where $Q=E_\mathit{jet}\theta_{c}=M_{jj}\theta_{c}/2$. All particles are treated as pions for Lorentz boosts. To evaluate possible biases that may originate from the particular choice of jet reconstruction algorithm, we compare results of the measurement using three different values of the parameter $R$ in the jet reconstruction algorithm (0.4, 0.7, 1.0). The resulting systematic uncertainty is $1\%$.

	We use full three-dimensional track reconstruction {\cite{CDFIItdr,ionim}}. Poorly reconstructed and spurious tracks are removed by a requirement on the quality of the track fit in the drift chamber $\chi^{2}_{COT}<6.0$ {\cite{ionim}}. Charged particles are required to have $p_{T}>0.3$ GeV/c. Requirements on the track impact parameter $d_{0}$, radius of conversion $R_{conv}$, and  $\left| \Delta z \right|=\left| z_{track}-z_{vertex} \right|$ are also applied (see {\cite{Correlations}} for details). These requirements are designed to ensure that the tracks originate at the primary vertex and are not produced by cosmic rays, multiple $p\bar p$ interactions within the same bunch crossing, $\gamma$ conversions, and $K^{0}$ and $\Lambda$ decays. The correction for the remaining fraction of secondary tracks is estimated by comparing the $k_{T}$ distribution in {\sc pythia} tune A at the charged hadron level and at the level of the detector simulation. It is found to be $\sim 3\%$ and is assigned as the systematic uncertainty associated with the remaining fraction of secondary tracks. In order to correct for tracks from the underlying event, we apply the following procedure. On an event-by-event basis, two complementary cones are positioned at the same polar angle with respect to the beamline as the original dijet axis but in the plane perpendicular to the dijet axis. We assume that cones formed in such a fashion collect statistically the same amount of background from the underlying event as the cones around the jet axis {\cite{SafonovPRD}}, and we subtract the $k_{T}$ distribution in complementary cones from the distribution in jet cones. 

\begin{figure}[h!]
	\includegraphics[width=2.52in]
	{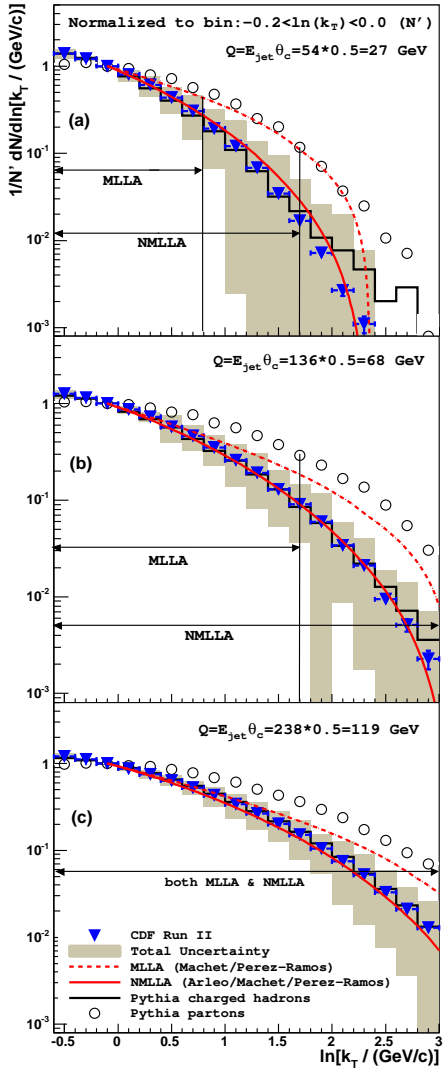}
	\caption{ The $k_{T}$ distribution of particles in the restricted cone of size $\theta_{c}=0.5$ around the jet axis in dijet mass bins with (a) $Q=27$, (b) $Q=68$, and (c) $Q=119$ GeV. The data are compared to the analytical MLLA and NMLLA predictions and to the predictions of the {\sc pythia} tune A Monte Carlo generator for partons and charged hadrons (shown as histograms). The distribution for partons is obtained using a parton shower cutoff value of $500$ MeV. Ranges of validity for MLLA and NMLLA predictions are shown by arrows.}
	\label{KtData}
	\end{figure}

	 Figure~{\ref{KtData}} shows the distributions in data corresponding to the dijet mass bins with $\left<Q\right>=27$ GeV ($95<M_{jj}<132$ GeV/$c^2$), $68$ GeV ($243<M_{jj}<323$ GeV/$c^2$), and $119$ GeV ($428<M_{jj}<563$ GeV/$c^2$). The distributions in the other five dijet mass bins are similar. The fraction of gluon jets in the sample, $f_{g}$, which is used to mix the theoretical prediction for quark and gluon jets, is obtained using {\sc pythia} tune A with the CTEQ5L parton distribution functions (PDFs) {\cite{CTEQ5L}}. $f_{g}$ decreases from 0.7 for $Q=19$ GeV to 0.2 for $Q=155$ GeV. The error bars correspond to the statistical uncertainty only, while the shaded area corresponds to the statistical and systematic uncertainties added in quadrature. The major source of systematic uncertainty in the measurement is the remaining fraction of secondary tracks. The systematic uncertainty due to the PDFs is evaluated by comparing results for the fraction of gluon jets obtained using the CTEQ5L and CTEQ6.1 \cite{CTEQ61} PDF sets, and is found to be negligible ($<~1\%$). The individual systematic uncertainties for results with different jet hardnesses are strongly correlated. 
 
 The solid line corresponds to the NMLLA theoretical curve \cite{RedamyKtNMLLA} for $Q_\mathit{eff}=230$ MeV, extracted from fits of inclusive momentum distributions \cite{SafonovPRD}. The dashed line corresponds to the MLLA theoretical curve calculated according to \cite{RedamyKtMLLA} for the same value of $Q_\mathit{eff}$.  The NMLLA predictions generally have a wider range of validity than the MLLA predictions. The NMLLA results for $Q_\mathit{eff}=230$ MeV provide an excellent description of the data over the entire range of particle $k_{T}$ and the dijet masses used in this measurement. The overall qualitative agreement between the data and the MLLA calculation \cite{RedamyKtMLLA} for $Q_\mathit{eff}=230$ MeV is very good within its range of validity. The extrapolation beyond the range (to higher $k_{T}$) fails to reproduce the data, predicting more particles than observed.

%	\begin{figure}[h!]
%	\includegraphics[width=2.55in]
%	{../Ktplots/mb2_CDF_MC_NMLLA_Log_Range_mod4.eps}
%	\includegraphics[width=2.55in]
%	{../Ktplots/mb5_CDF_MC_NMLLA_Log_Range_mod4.eps}
%	\includegraphics[width=2.55in]
%	{../Ktplots/mb7_CDF_MC_NMLLA_Log_Range_mod4.eps}
%	\caption{ The $k_{T}$ distribution of particles in the restricted cone of size $\theta_{c}=0.5$ around the jet axis in dijet mass bins with (a) $Q=27$, (b) $Q=68$, and (c) $Q=119$ GeV. The data are compared to the analytical MLLA and NMLLA predictions and to the predictions of the {\sc pythia} tune A Monte Carlo generator for partons and charged hadrons (shown as histograms). Ranges of validity for MLLA and NMLLA predictions are shown by arrows.}
%	\label{KtData}
%	\end{figure}

	We also compare the $k_{T}$ distribution of charged particles in data to predictions of the {\sc pythia} tune A and {\sc herwig} 6.5 Monte Carlo generators. Predictions of the Monte Carlo generators for final stable particles are in agreement with each other and with results obtained in data. Figure~{\ref{KtData}} shows distributions in data compared to {\sc pythia} tune A at the parton and the final stable particle levels. The distribution for partons is obtained using a parton shower cutoff value of $500$ MeV, the lowest possible setting in the generator. The qualitative agreement between the NMLLA predictions and charged hadrons from {\sc pythia} tune A is found to be fairly good and is due to the tunings of the hadronization parameters in {\sc pythia} tune A,  while the distribution at the parton level shows significant deviations. The {\sc herwig}~6.5 predictions at the level of final stable particles are similar to those of {\sc pythia} tune A.

	In summary, we have measured the transverse momenta of particles with respect to the jet axis for a wide range of dijet invariant masses, 66--737 GeV/c$^{2}$. The data are compared to calculations using the modified leading log and next-to-modified leading log approximations. Within the range of their validity, the next-to-modified leading log approximation calculations provide an excellent description of trends seen in the data over the entire range of dijet masses. This indicates that hadronization effects are small and provides further support for the hypothesis of local parton-hadron duality. The modified leading log approximation predictions qualitatively show the same trends; however, the quantitative disagreement with the data is significant in this case, indicating the importance of the next-to-modified leading log approximation corrections. 
	The authors are very grateful to R.~Perez-Ramos, F.~Arleo, and B.~Machet for collaborative work and to Yu.~Dokshitzer for a number of very fruitful discussions. We thank the Fermilab staff and the technical staffs of the participating institutions for their vital contributions. This work was supported by the U.S. Department of Energy and National Science Foundation; the Italian Istituto Nazionale di Fisica Nucleare; the Ministry of Education, Culture, Sports, Science and Technology of Japan; the Natural Sciences and Engineering Research Council of Canada; the National Science Council of the Republic of China; the Swiss National Science Foundation; the A.P. Sloan Foundation; the Bundesministerium f\"ur Bildung und Forschung, Germany; the Korean Science and Engineering Foundation and the Korean Research Foundation; the Science and Technology Facilities Council and the Royal Society, UK; the Institut National de Physique Nucleaire et Physique des Particules/CNRS; the Russian Foundation for Basic Research; the Ministerio de Ciencia e Innovaci\'{o}n, and Programa Consolider-Ingenio 2010, Spain; the Slovak R\&D Agency; and the Academy of Finland.

\end{document}

%% file: July_2008_Authors_Visitors.tex
\affiliation{Institute of Physics, Academia Sinica, Taipei, Taiwan 11529, Republic of China} 
\affiliation{Argonne National Laboratory, Argonne, Illinois 60439} 
\affiliation{University of Athens, 157 71 Athens, Greece} 
\affiliation{Institut de Fisica d'Altes Energies, Universitat Autonoma de Barcelona, E-08193, Bellaterra (Barcelona), Spain} 
\affiliation{Baylor University, Waco, Texas  76798} 
\affiliation{Istituto Nazionale di Fisica Nucleare Bologna, $^v$University of Bologna, I-40127 Bologna, Italy} 
\affiliation{Brandeis University, Waltham, Massachusetts 02254} 
\affiliation{University of California, Davis, Davis, California  95616} 
\affiliation{University of California, Los Angeles, Los Angeles, California  90024} 
\affiliation{University of California, San Diego, La Jolla, California  92093} 
\affiliation{University of California, Santa Barbara, Santa Barbara, California 93106} 
\affiliation{Instituto de Fisica de Cantabria, CSIC-University of Cantabria, 39005 Santander, Spain} 
\affiliation{Carnegie Mellon University, Pittsburgh, PA  15213} 
\affiliation{Enrico Fermi Institute, University of Chicago, Chicago, Illinois 60637} 
\affiliation{Comenius University, 842 48 Bratislava, Slovakia; Institute of Experimental Physics, 040 01 Kosice, Slovakia} 
\affiliation{Joint Institute for Nuclear Research, RU-141980 Dubna, Russia} 
\affiliation{Duke University, Durham, North Carolina  27708} 
\affiliation{Fermi National Accelerator Laboratory, Batavia, Illinois 60510} 
\affiliation{University of Florida, Gainesville, Florida  32611} 
\affiliation{Laboratori Nazionali di Frascati, Istituto Nazionale di Fisica Nucleare, I-00044 Frascati, Italy} 
\affiliation{University of Geneva, CH-1211 Geneva 4, Switzerland} 
\affiliation{Glasgow University, Glasgow G12 8QQ, United Kingdom} 
\affiliation{Harvard University, Cambridge, Massachusetts 02138} 
\affiliation{Division of High Energy Physics, Department of Physics, University of Helsinki and Helsinki Institute of Physics, FIN-00014, Helsinki, Finland} 
\affiliation{University of Illinois, Urbana, Illinois 61801} 
\affiliation{The Johns Hopkins University, Baltimore, Maryland 21218} 
\affiliation{Institut f\"{u}r Experimentelle Kernphysik, Universit\"{a}t Karlsruhe, 76128 Karlsruhe, Germany} 
\affiliation{Center for High Energy Physics: Kyungpook National University, Daegu 702-701, Korea; Seoul National University, Seoul 151-742, Korea; Sungkyunkwan University, Suwon 440-746, Korea; Korea Institute of Science and Technology Information, Daejeon, 305-806, Korea; Chonnam National University, Gwangju, 500-757, Korea} 
\affiliation{Ernest Orlando Lawrence Berkeley National Laboratory, Berkeley, California 94720} 
\affiliation{University of Liverpool, Liverpool L69 7ZE, United Kingdom} 
\affiliation{University College London, London WC1E 6BT, United Kingdom} 
\affiliation{Centro de Investigaciones Energeticas Medioambientales y Tecnologicas, E-28040 Madrid, Spain} 
\affiliation{Massachusetts Institute of Technology, Cambridge, Massachusetts  02139} 
\affiliation{Institute of Particle Physics: McGill University, Montr\'{e}al, Qu\'{e}bec, Canada H3A~2T8; Simon Fraser University, Burnaby, British Columbia, Canada V5A~1S6; University of Toronto, Toronto, Ontario, Canada M5S~1A7; and TRIUMF, Vancouver, British Columbia, Canada V6T~2A3} 
\affiliation{University of Michigan, Ann Arbor, Michigan 48109} 
\affiliation{Michigan State University, East Lansing, Michigan  48824}
\affiliation{Institution for Theoretical and Experimental Physics, ITEP, Moscow 117259, Russia} 
\affiliation{University of New Mexico, Albuquerque, New Mexico 87131} 
\affiliation{Northwestern University, Evanston, Illinois  60208} 
\affiliation{The Ohio State University, Columbus, Ohio  43210} 
\affiliation{Okayama University, Okayama 700-8530, Japan} 
\affiliation{Osaka City University, Osaka 588, Japan} 
\affiliation{University of Oxford, Oxford OX1 3RH, United Kingdom} 
\affiliation{Istituto Nazionale di Fisica Nucleare, Sezione di Padova-Trento, $^w$University of Padova, I-35131 Padova, Italy} 
\affiliation{LPNHE, Universite Pierre et Marie Curie/IN2P3-CNRS, UMR7585, Paris, F-75252 France} 
\affiliation{University of Pennsylvania, Philadelphia, Pennsylvania 19104}
\affiliation{Istituto Nazionale di Fisica Nucleare Pisa, $^x$University of Pisa, $^y$University of Siena and $^z$Scuola Normale Superiore, I-56127 Pisa, Italy} 
\affiliation{University of Pittsburgh, Pittsburgh, Pennsylvania 15260} 
\affiliation{Purdue University, West Lafayette, Indiana 47907} 
\affiliation{University of Rochester, Rochester, New York 14627} 
\affiliation{The Rockefeller University, New York, New York 10021} 
\affiliation{Istituto Nazionale di Fisica Nucleare, Sezione di Roma 1, $^{aa}$Sapienza Universit\`{a} di Roma, I-00185 Roma, Italy} 

\affiliation{Rutgers University, Piscataway, New Jersey 08855} 
\affiliation{Texas A\&M University, College Station, Texas 77843} 
\affiliation{Istituto Nazionale di Fisica Nucleare Trieste/Udine, $^{bb}$University of Trieste/Udine, Italy} 
\affiliation{University of Tsukuba, Tsukuba, Ibaraki 305, Japan} 
\affiliation{Tufts University, Medford, Massachusetts 02155} 
\affiliation{Waseda University, Tokyo 169, Japan} 
\affiliation{Wayne State University, Detroit, Michigan  48201} 
\affiliation{University of Wisconsin, Madison, Wisconsin 53706} 
\affiliation{Yale University, New Haven, Connecticut 06520} 
\author{T.~Aaltonen}
\affiliation{Division of High Energy Physics, Department of Physics, University of Helsinki and Helsinki Institute of Physics, FIN-00014, Helsinki, Finland}
\author{J.~Adelman}
\affiliation{Enrico Fermi Institute, University of Chicago, Chicago, Illinois 60637}
\author{T.~Akimoto}
\affiliation{University of Tsukuba, Tsukuba, Ibaraki 305, Japan}
\author{B.~\'{A}lvarez~Gonz\'{a}lez}
\affiliation{Instituto de Fisica de Cantabria, CSIC-University of Cantabria, 39005 Santander, Spain}
\author{S.~Amerio$^w$}
\affiliation{Istituto Nazionale di Fisica Nucleare, Sezione di Padova-Trento, $^w$University of Padova, I-35131 Padova, Italy} 

\author{D.~Amidei}
\affiliation{University of Michigan, Ann Arbor, Michigan 48109}
\author{A.~Anastassov}
\affiliation{Northwestern University, Evanston, Illinois  60208}
\author{A.~Annovi}
\affiliation{Laboratori Nazionali di Frascati, Istituto Nazionale di Fisica Nucleare, I-00044 Frascati, Italy}
\author{J.~Antos}
\affiliation{Comenius University, 842 48 Bratislava, Slovakia; Institute of Experimental Physics, 040 01 Kosice, Slovakia}
\author{G.~Apollinari}
\affiliation{Fermi National Accelerator Laboratory, Batavia, Illinois 60510}
% ...

\collaboration{CDF Collaboration\footnote{With visitors from $^a$University of Massachusetts Amherst, Amherst, Massachusetts 01003,
$^b$Universiteit Antwerpen, B-2610 Antwerp, Belgium, 
$^c$University of Bristol, Bristol BS8 1TL, United Kingdom,
$^d$Chinese Academy of Sciences, Beijing 100864, China, 
$^e$Istituto Nazionale di Fisica Nucleare, Sezione di Cagliari, 09042 Monserrato (Cagliari), Italy,
$^f$University of California Irvine, Irvine, CA  92697, 
$^g$University of California Santa Cruz, Santa Cruz, CA  95064, 
$^h$Cornell University, Ithaca, NY  14853, 
$^i$University of Cyprus, Nicosia CY-1678, Cyprus, 
$^j$University College Dublin, Dublin 4, Ireland,
$^k$Royal Society of Edinburgh/Scottish Executive Support Research Fellow,
$^l$University of Edinburgh, Edinburgh EH9 3JZ, United Kingdom, 
$^m$Universidad Iberoamericana, Mexico D.F., Mexico,
$^n$Queen Mary, University of London, London, E1 4NS, England,
$^o$University of Manchester, Manchester M13 9PL, England, 
$^p$Nagasaki Institute of Applied Science, Nagasaki, Japan, 
$^q$University of Notre Dame, Notre Dame, IN 46556,
$^r$University de Oviedo, E-33007 Oviedo, Spain, 
$^s$Texas Tech University, Lubbock, TX  79409, 
$^t$IFIC(CSIC-Universitat de Valencia), 46071 Valencia, Spain,
$^u$University of Virginia, Charlottesville, VA  22904,
$^{cc}$On leave from J.~Stefan Institute, Ljubljana, Slovenia, 
}}
\noaffiliation

%% file: prl_Kt_collab3_1.bbl
\begin{references}
\bibitem{Et}  $E_{T}$ here is defined as energy multiplied by $\sin(\theta)$ where $\theta$ is the polar angle 
               with respect to the beam axis.
\end{references}
